\documentclass[11pt]{article}

\def\mytitle#1{\setcounter{equation}{0}
\setcounter{footnote}{0}
\begin{flushleft}\Large\textbf{#1}\end{flushleft}
\vspace{0.25cm}}
\def\myname#1{\leftline{{\large #1}}\vspace{-0.13cm}}
\def\myplace#1#2{\small\begin{flushleft}\textit{#1}\\
\texttt{#2}\end{flushleft}}

\usepackage{graphicx}% Include figure files
\usepackage{amsmath, amssymb, graphics}
\begin{document}

\mytitle{Reconstruction of \textit{f(R)} gravity with ordinary and
Entropy-corrected \textit{(m,n)}-type Holographic dark energy
model}

\vskip0.2cm \myname{Prabir Rudra\footnote{prudra.math@gmail.com}}
\myplace{Department of Mathematics, Asutosh College, Kolkata-700
026, India.}{}

\begin{abstract}
In this assignment we will present a reconstruction scheme between
\textit{f(R)} gravity with ordinary and entropy corrected
$(m,n)$-type holographic dark energy. The correspondence is
established and expressions for the reconstructed $f(R)$ models
are determined. To study the evolution of the reconstructed models
plots are generated. The stability of the calculated models are
also investigated using the squared speed of sound in the
background of the reconstructed gravities.
\end{abstract}

\vspace{5mm}


\vspace{5mm}

{\it Pacs. No.: 98.80.-k, 98.80.Es, 95.35.+d, 95.36.+x}\\

\vspace{5mm}

\section{Introduction}
The acceleration of the universe, as confirmed by various
cosmological observations such as Ia type supernovae, CMB, WMAP,
etc. has been the greatest cosmological discovery over the last 50
years \cite{Perlmutter, Riess, Spergel}. In the background of the
classical Newton's theory of gravitation, this seemed to be quite
an unusual and unexpected phenomenon. The reason behind this can
be attributed to the fact that the classical theory does not give
us any idea regarding the repulsive nature of gravity which is
quite evident from the accelerated expansion of the universe. As a
possible explanation to the phenomenon, scientists resorted to a
mysterious negative pressure component named the dark energy.
Instead of competition from other counterparts such as the theory
of modified gravity, dark energy has gained popularity of the last
decade. Nevertheless, dark energy and modified gravity theory
complements each other as is evident from literature \cite{Bamba1,
Bamba2, Nojiri1}. Observations suggest that almost two-thirds of
the total energy of the universe is contributed by this mysterious
form of energy. The rest of the portion is occupied by dark matter
with a little contribution from baryonic matter \cite{Bennett}.

In the quest of a quantum gravity theory, Fischler and Suskind
\cite{Fisher1, Susskind1} proposed the holographic principle. In
accordance with the holographic principle the the dark energy in
the context of quantum gravity is known as the holographic dark
energy (HDE). Cohen et al. \cite{Cohen1} proposed an enlargement
relationship between the infra-red (IR) and ultraviolet cut-offs.
The relationship is basically due to the limitation set by the
formation of a black hole, which establishes an upper limit for
the vacuum energy, $L^{3}\rho_{v}\leq LM_{P}^{2}$, where
$\rho_{v}$ is the HDE density related to the UV cut-off, $L$ is
the IR cut-off and $M_{P}$ is the reduced Planck mass. Over the
years various dark energy that have been developed which remain
plagued with different cosmological problems. The notable ones
being the cosmic coincidence problem, fine tuning problem, etc. It
must be noted that HDE may provide simultaneous natural solutions
to both dark energy and cosmological problems \cite{Li1}.
Precisely, it gives excellent solutions to the coincidence problem
\cite{Pavon1} and the phantom crossing \cite{Wang1}. The novel
feature of the model is that it has reasonable agreement with the
astrophysical data of CMB, SNeIa and galaxy redshift surveys
\cite{Wei1}. As a result scientists have been inclined to extend
the idea of HDE dark energy via various prescriptions. Extensions
using different cut-offs is done in the ref. \cite{Granda1} and
various entropy corrections have been considered in
\cite{Setare1}. The holographic scale denoted by $L$ has a direct
relation with the future event horizon \cite{Li1}, the conformal
age of the universe \cite{Wei2} or the Ricci scalar of the
universe \cite{Gao1}. The model taking into account the conformal
age of the universe is widely known as new agegraphic dark energy
(NADE) model in literature. Recently an extension to this idea has
been introduced by the name of $(m, n)-$ type HDE \cite{Ling1},
where $m$ and $n$ are the parameters associated with the chosen IR
cut-off given by,
\begin{equation}
L=\frac{1}{a^{m}(t)}\int_{0}^{t}a^{n}(t')dt'
\end{equation}

If we consider the consistency of a HDE model with a
conformal-like age of the universe as the scale $L$ with the past
inflationary phase, perhaps the direct physical motivation of
proposing such characteristic scales for the $(m, n)$-type
holographic dark energy model is still obscure. In spite of such
theoretical difficulty the model has some significant advantages
that propelled its survival. The model generalizes the theory with
noteworthy improvements. Moreover the parameters $(m, n)$ opens up
space to fit observational data in the background of a sound
theoretical framework.

It is seen that various cosmological scenarios can be achieved by
giving some specific values to $(m, n)$. The transition of the
Equation of state (EoS) across the phantom divide can easily be
achieved for some particular values of $(m, n)$ even without the
introduction of any interaction between dark energy and dark
matter. Similarly for suitable values of $(m, n)$, all the
agegraphic-like dark energy models can be recovered. For age-like
holographic models, when $m = n$ it seems that dark energy behaves
as the dominant ingredient in the early epochs of the universe.
This implies the unification of dark energy with dark matter. But
we need to introduce some mechanism that can differentiate dark
energy from the dark matter state in the early epoch and help us
to identify the dominating nature of dark energy in the late
universe, which will ultimately drive the acceleration of the
universe. An appropriate interaction between dark energy and dark
matter serves the purpose. The stability of the model in the dark
energy dominated era has already been confirmed. In ref.
\cite{Huang} it is seen that, best-fit analysis with observational
data indicates that this model with $m = n + 1$ and small $m$ is
more favoured and physically more viable.

From literature, we know that a suitable alternative to the
concept of dark energy in justifying the late cosmic acceleration
lies in the modification of the theory of gravity, i.e, modifying
the Einstein-Hilbert action. A popular way is to use an analytic
function $f=f(R)$ in place of $R$ in the action, where $R$ is the
Ricci scalar \cite{Capo, Nojiri2}. The modification basically
deals with the geometry of space-time and is responsible for
varying its curvature. In this connection it must be noted that
$f(R) \propto R^{2}$ kind of theories could successfully generate
inflationary scenario for the early universe \cite{Staro}. Since
the curvature $R$ decreases as the universe evolves, any inverse
power of $R$ in the expression of $f(R)$ will generate a late time
accelerating universe. Detailed description of $f(R)$ theories can
be widely found in literature \cite{Soti, Felice}. Although most
of the works in literature deal with the present day acceleration,
yet recently some works can be found where the authors have
considered the smooth transition from decelerated to an
accelerated regime \cite{Nojiri1}. Moreover it has been a
challenge to realize both the early time inflation and the late
time acceleration from the standard $f(R)$ gravity models. However
in this regard some light has been thrown in ref. \cite{Cognola,
Elizalde}. Recently reconstruction scheme has been a very
interesting topic of study in cosmology. Reconstruction scheme has
been studied using $f(R)$ gravity on many occasions by different
authors \cite{Nojiri3}. In ref. \cite{Rahul} the authors have
discussed a reconstruction scheme of $f(G)$ gravity with $(m,
n)$-type and entropy corrected $(m, n)$-type HDE models. Motivated
by this, we set to develop a reconstruction scheme between $f(R)$
gravity and two types of HDE models, namely the $(m, n)$-type HDE
model and entropy corrected $(m, n)$-type HDE (ECHDE) model in
this assignment.

The Einstein field equations in the FRW model in the background of
$f(R)$ gravity are discussed in section $2$. Reconstruction scheme
with $(m, n)$-type HDE and entropy corrected $(m, n)$-type HDE is
studied in section $2.1$ and $2.2$ respectively. The stability of
the reconstructed models is discussed in section $3$. Finally the
paper ends with a conclusion in section $4$.

\section{Reconstruction in \textit{f(R)} gravity}
The generalized Einstein-Hilbert action for $f(R)$ gravity is
given by,
\begin{equation}
S=\int \left[\frac{1}{16\pi G}f(R)+\cal{L}\right]\sqrt{-g}d^{4}x
\end{equation}
where $R$ is replaced by $f(R)$ in the Einstein-Hilbert action.
Here $f(R)$ is an analytic function of $R$ and $\cal{L}$ is the
usual matter field Lagrangian. The corresponding field equations
in FRW space-time is given by,
\begin{equation}
3\frac{\dot{a}^{2}}{a^{2}}=\frac{\rho_{m}}{f_{R}}+\frac{1}{f_{R}}\left[\frac{1}{2}\left(f(R)-Rf_{R}\right)-3\dot{R}f_{RR}\frac{\dot{a}}{a}\right]
\end{equation}
and
\begin{equation}
2\frac{\ddot{a}}{a}+\frac{\dot{a}^{2}}{a^{2}}=-\frac{1}{f_{R}}\left[\ddot{R}f_{RR}+\dot{R}^{2}f_{RRR}+2\dot{R}f_{RR}\frac{\dot{a}}{a}-\frac{1}{2}\left(f(R)-Rf_{R}\right)\right]
\end{equation}
where dots and primes denote derivative with respect to cosmic
time and $R$ respectively. Here $\rho_{m}$ is the dark matter
density, and the corresponding pressure $p_{m}$ is taken to be
zero. So we can write the contribution by curvature to density and
pressure as,
\begin{equation}
\rho_{R}=\left[\frac{1}{2}\left(f(R)-Rf_{R}\right)-3\dot{R}f_{RR}\frac{\dot{a}}{a}\right]=\frac{1}{2}\left(f(R)-Rf_{R}\right)+18\left(\ddot{H}+4H\dot{H}\right)Hf_{RR}
\end{equation}
and
\begin{equation}
p_{R}=\left[\ddot{R}f_{RR}+\dot{R}^{2}f_{RRR}+2\dot{R}f_{RR}\frac{\dot{a}}{a}-\frac{1}{2}\left(f(R)-Rf_{R}\right)\right]
\end{equation}
where $R=-6\left(\dot{H}+2H^{2}\right)$ and
$\dot{R}=-6\left(\ddot{H}+4H\dot{H}\right)$.

\subsection{Reconstruction with (m,n)-type HDE model}
The energy density of the $(m, n)$-type HDE model as given by ref.
\cite{Ling1} is,
\begin{equation}
\rho_{v}=\frac{3b^{2}}{L^{2}}
\end{equation}
where $b$ is a constant and $L$ is the generalized IR cut-off
defined as,
\begin{equation}
L=\frac{1}{a^{m}(t)}\int_{0}^{t}a^{n}(t')dt'
\end{equation}
where $m$ and $n$ are constants.

A simple calculation gives
\begin{equation}
\dot{L}=-mHL+a^{n-m}(t)
\end{equation}
Now we establish the correspondence by setting
$\rho_{v}=\rho_{R}$. We get,
\begin{equation}
\frac{1}{2}\left(f(R)-Rf_{R}\right)+18\left(\ddot{H}+4H\dot{H}\right)Hf_{RR}=
\frac{b^{2}}{L^{2}}\left(\frac{2a^{n-m}}{HL}-2m-3\right)
\end{equation}
where $f_{R}$ and $f_{RR}$ are respectively the first and second
order derivatives of $f(R)$ with respect to $R$. We consider the
scale factor as the power law form of time $a(t)=a_{0}t^{p}$,
where $p$ is a positive constant.

The above is a differential equation of second order. Solving this
we get,

$$f(R)=\frac{A}{6}a_{0}^{-3n}\left(6^{p/2}a_{0}\left(\frac{B}{R}\right)^{p/2}\right)^{-m}\left(\frac{B}{R}\right)
^{3(m-n)p/2}\left[\exp(-p/2)\left(-4a_{0}^{3m}b^{2}\exp(p/2)\sqrt{B/p}\left(1+np\right)^{3}\times\right.\right.$$

$$\left.\left.\left(6^{p/2}a_{0}\left(B/R\right)^{p/2}\right)^{n}R+2a_{0}^{2m+n}b^{2}\exp(p/2)\sqrt{Bp}
\left(2A^{-1/3}m\left(1+np\right)^{2}+3A^{-1/3}\left(1+2np+n^{2}p^{2}\right)\right)\times\right.\right.$$

$$\left.\left.\left(6^{p/2}a_{0}\frac{B}{R}^{p/2}\right)^{m}\left(\frac{B}{R}\right)^{(n-m)p/2}R-6a_{0}^{3n}p^{2}
\left(6^{p/2}a_{0}\frac{B}{R}^{p/2}\right)^{m}\left(\frac{B}{R}\right)^{3(n-m)p/2}C_{1}\frac{\exp(p/2)}{A}
\left(1-2p\right)^{2}\right.\right.$$

$$\left.\left.-\exp(1/4)\sqrt{B/p}~C_{2}\left(-\frac{1}{A}+\frac{2p}{A}\right)\right)+\frac{3C_{2}}{A}a_{0}^{3n}\sqrt{\pi}
\left(6^{p/2}a_{0}\left(\frac{B}{R}\right)^{p/2}\right)^{m}\left(\frac{B}{R}\right)^{2+3(n-m)p/2}R^{2}\times\right.$$
\begin{equation}
\left.Erfi\left(\frac{1}{2}\sqrt{\frac{B}{p}}\right)\right]
\end{equation}
where
$A=6^{\frac{3}{2}\left(m-n\right)p}$,~~~$B=p\left(1-2p\right)$~~
and~~ $Erfi$~~ represents the imaginary error function. $C_{1}$
and $C_{2}$ are constants of integration.\\

\begin{figure}
\vspace{3mm}
~~~~~~~~~~~~~~~~~~~~~~~~~~~~~\includegraphics[height=3in, width=3in]{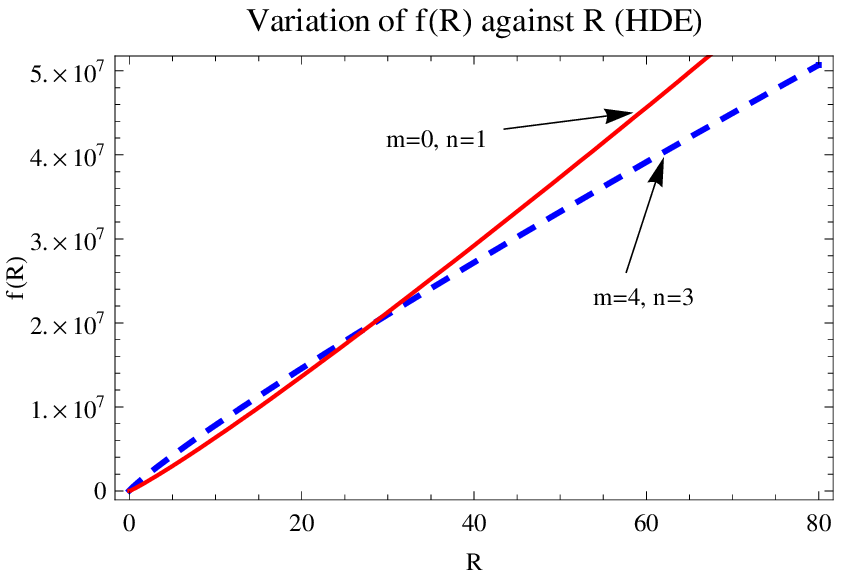}~~~~~~~\\
\vspace{1mm}
~~~~~~~~~~~~~~~~~~~~~~~~~~~~~~~~~~~~~~~~~~~~~~~~~~~~~~~~~~~~~Fig.1~~~~~~~~~~~~~~~~~~\\
\vspace{3mm} \textit{\textbf{Fig 1} shows the evolution of $f(R)$
gravity against the Ricci scalar $R$ for HDE dark energy for
different values of $m$ and $n$. The other parameters are
considered as $p=0.1, a_{0}=1, b=1, C_{1}=2, C_{2}=3$.}
\end{figure}

Evolution of $f(R)$ against $R$ for the HDE model is plotted in
fig.1 for two different sets of parametric values ($m=4$, $n=3$
and $m=0$, $n=1$). From literature \cite{Huang} the best-fit
analysis indicates that the models with $m=n+1$ and relatively
small values of $m$ is more favoured. This includes the cases of
$(m, n)=(1, 0),(2, 1),(3, 2),(4, 3),$ etc. For the purpose of
study, here we choose two different pairs of parametric values,
one following the above rule ($m=4$, $n=3$) and the other
violating it ($m=0$, $n=1$). The idea is to compare the results in
the two cases. From the figure it is seen that $f(R)$ increases
with $R$ for both set of values of $m$ and $n$. For $m=4, n=3$, we
see that $f(R)$ varies linearly with $R$. But for $m=0, n=1$,
$f(R)$ follows a non-linear relation in $R$. The two curves
intersect at around $R=30$.

\subsection{Reconstruction with entropy corrected (m,n)-type HDE model}
Considering the corrected entropy-area relation with derivation of
HDE, we can obtain the energy density of the entropy-corrected (m,
n)-type HDE as,
\begin{equation}
\rho_{V}=\frac{3c^{2}}{L^{2}}+\frac{\xi}{L^{4}}\log{L^{2}}+\frac{\zeta}{L^{4}}
\end{equation}
where $c$, $\xi$ and $\zeta$ are dimensionless constant and IR
cutoff as considered in eqn.(1). Here also to develop the
reconstruction scheme we have to consider $\rho_{R}=\rho_{V}$,
which gives,
\begin{equation}
\frac{1}{2}\left(f(R)-Rf_{R}\right)+18\left(\ddot{H}+4H\dot{H}\right)Hf_{RR}=
\frac{3c^{2}}{L^{2}}+\frac{\xi}{L^{4}}\log{L^{2}}+\frac{\zeta}{L^{4}}
\end{equation}
Now we consider a power law form of scale factor $a=a_{0}t^{p}$, p
being a positive constant, as considered in the previous case.
Solving the above differential equation we get,

$$f(R)=\frac{1}{36}\left[36C_{3}E+\frac{18e^{-p/2}C_{4}}{E}\left\{2e^{1/4}E-E^{2}e^{p/2}\sqrt{\pi}
Erfi(E/2)\right\}+\frac{1}{R}2^{1-2np}9^{-np}a_{0}^{2(m-2n)}\left(1+np\right)^{2}\times\right.$$

$$\left.\left(pE^{2}/R\right)^{2(-1+p(m-n))}\left(2^{1+(m+n)p}3^{2+(m+n)p}a_{0}^{2n}c^{2}E^{2}p
\left(pE^{2}/R\right)^{(n-m)p}+a_{0}^{2m}F\zeta R\right.\right.$$

\begin{equation}
\left.\left.+a_{0}^{2m}F\xi R
\log{\frac{6^{1-mp+np}a_{0}^{2(n-m)}\left(E^{2}p/R\right)^{1+p(n-m)}}{\left(1+np\right)^{2}}}\right)\right]
\end{equation}
where $E=\sqrt{1-2p}$,
~~$F=36^{mp}+2^{1+2mp}9^{mp}np+36^{mp}n^{2}p^{2}$. $C_{3}$ and
$C_{4}$ are constants of integration.

\vspace{5mm}
\begin{figure}
\vspace{3mm}
~~~~~~~~~~~~~~~~~~~~~~~~~~~~~\includegraphics[height=3in, width=3in]{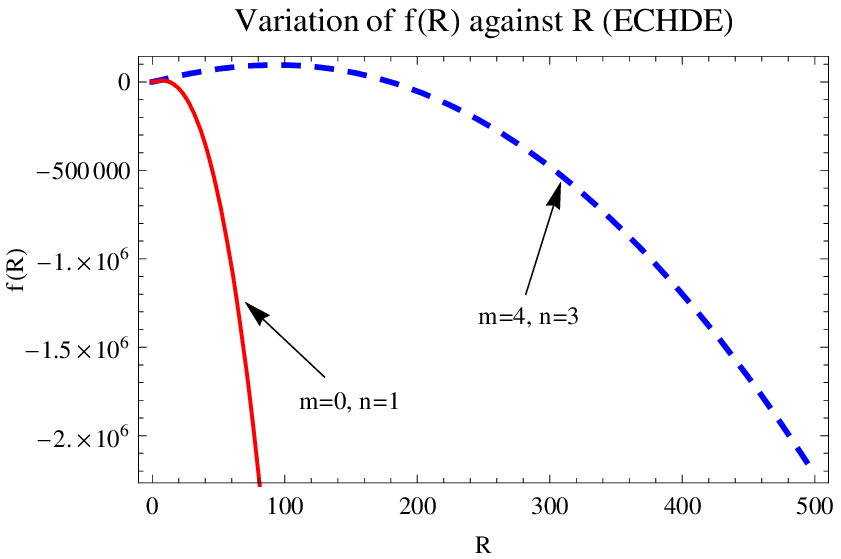}~~~~~~~\\
\vspace{1mm}
~~~~~~~~~~~~~~~~~~~~~~~~~~~~~~~~~~~~~~~~~~~~~~~~~~~~~~~~~~~~~Fig.2~~~~~~~~~~~~~~~~~~\\
\vspace{3mm} \textit{\textbf{Fig 2} shows the evolution of $f(R)$
gravity against the Ricci scalar $R$ for ECHDE dark energy for
different values of $m$ and $n$. The other parameters are
considered as $p=0.1, a_{0}=1, c=10, C_{3}=1, C_{4}=1, \xi=1,
\zeta=5$.}
\end{figure}
\vspace{3mm}

Evolution of $f(R)$ against $R$ for the ECHDE model is plotted in
fig.2 for two different sets of parametric values ($m=4$, $n=3$
and $m=0$, $n=1$). From the figure it is seen that $f(R)$ decays
with $R$ for both set of values of $m$ and $n$. For $m=0, n=1$,
the curve drops steeply as compared to the curve corresponding to
the values $m=4, n=3$, both following non-linear evolution.

\section{Stability Analysis}
The squared speed of sound $v_{s}^{2}=\frac{\dot{p}}{\dot{\rho}}$
is the parameter that helps us to study the stability of the
background evolution of a particular gravity model. The positive
value of $v_{s}^{2}$ characterizes stability of the corresponding
model \cite{Sharif, Kim}. The new HDE model in an interacting
scenario produces a negative $v_{s}^{2}$, thus exhibiting
classical instability \cite{Sharif}. Choosing the future event
horizon as IR cut-off, the squared speed of sound is found to stay
in the negative level for HDE, whereas for Chaplygin gas and
tachyon model, we get non-negative values of $v_{s}^{2}$
\cite{Myung}. In ref. \cite{Kim}, it is observed that the perfect
fluid for agegraphic dark energy is classically unstable.

In this study we have calculated $v_{s}^{2}$ and plotted it
against cosmic time $t$ for both the reconstructed $f(R)$ models.
In fig.3, for HDE, we see that for $m=4$ and $n=3$, $v_{s}^{2}$
stays in the negative level as the universe evolves with time. The
ordinate line corresponding to $t=9$ is un-physical and so we do
not include it in our discussion. So for this case we witness a
classical instability of the model. In fig.4, we have generated
the $v_{s}^{2}$ vs $t$ plot for parametric values $m=0$, $n=1$.
The asymptotic nature of the curve above the $t$ axis implies that
the squared speed of sound remains in the positive level
throughout the evolution. So here we see that the model is stable.
In figs.5 and 6, plots of have been generated for $v_{s}^{2}$ for
ECHDE. In fig.5, it is seen that for $m=4$ and $n=3$, $v_{s}^{2}$
stays in the negative level for greater part of the evolution. It
is to be noted that the curve undergoes a transition from positive
to negative level at some time relatively in early universe. This
is a contrast to what we obtained in case of HDE for the
corresponding case. Unlike that case, here we get a stable model
in the early universe, but as the universe evolves, the model
becomes unstable. Here we speculate that the late cosmic
acceleration plays a big role in destabilizing the evolution of
the universe, which is expected from general notion of dynamics.
In fig.6, for $m=0$, $n=1$, we see that the model exhibits a
perfectly stable nature. From the above discussion, we see that
the stability of our models is parameter dependent.

In \cite{Rahul}, the correspondence have been obtained between
$f(G)$ gravity and dark energy models HDE and ECHDE. For both the
HDE and ECHDE case, we have obtained a complete contrast with the
study in ref \cite{Rahul}. From the previous studies of
reconstruction scheme \cite{Sharif, Myung} using HDE it was seen
that classical instability characterized the models. So if we
consider them and compare them with this assignment, it is quite
obvious that $m=4$, $n=3$ is the better choice of parameters,
which follows the $m=n+1$ rule \cite{Huang}. This set of
parameters also give an unstable model for ECHDE.

\begin{figure}
\vspace{3mm}
~\includegraphics[height=3in, width=3in]{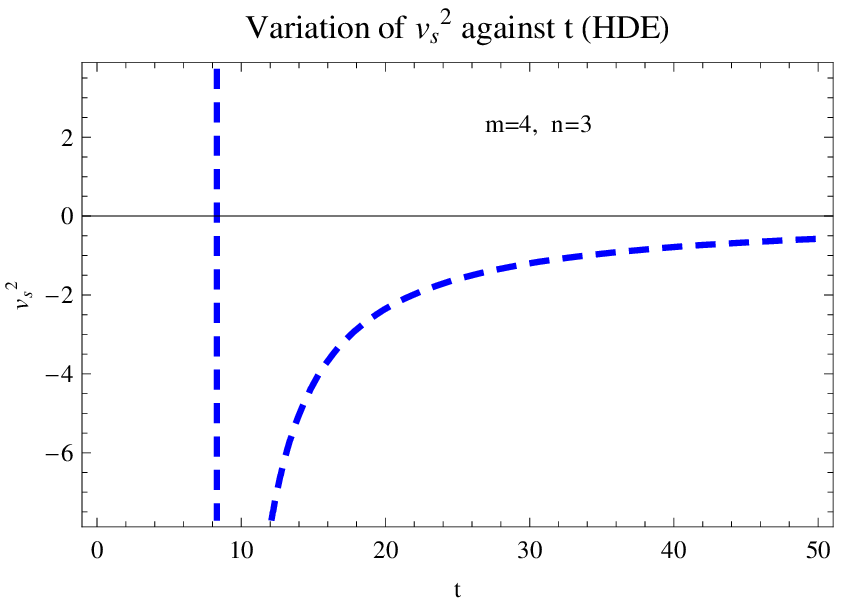}~~~~~~~~~\includegraphics[height=3in, width=3in]{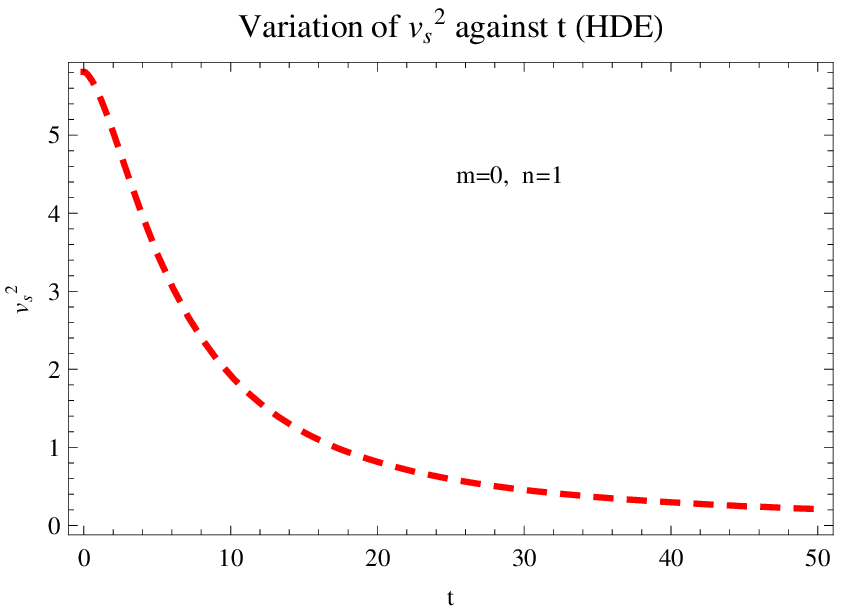}~~~~~~~\\
\vspace{1mm}
~~~~~~~~~~~~~~~~~~~~~~~~~~~~~~~~~Fig.3~~~~~~~~~~~~~~~~~~~~~~~~~~~~~~~~~~~~~~~~~~~~~~~~~~~~~~~~~~~Fig.4~~~~~~~~~~~~~~~~~\\
\vspace{3mm} \textit{\textbf{Fig 3, 4} shows the variation of
squared speed of sound $v_{s}^{2}$ against time $t$ for HDE dark
energy for different values of $m$ and $n$.}

\vspace{3mm}
~~~~\includegraphics[height=3in, width=3in]{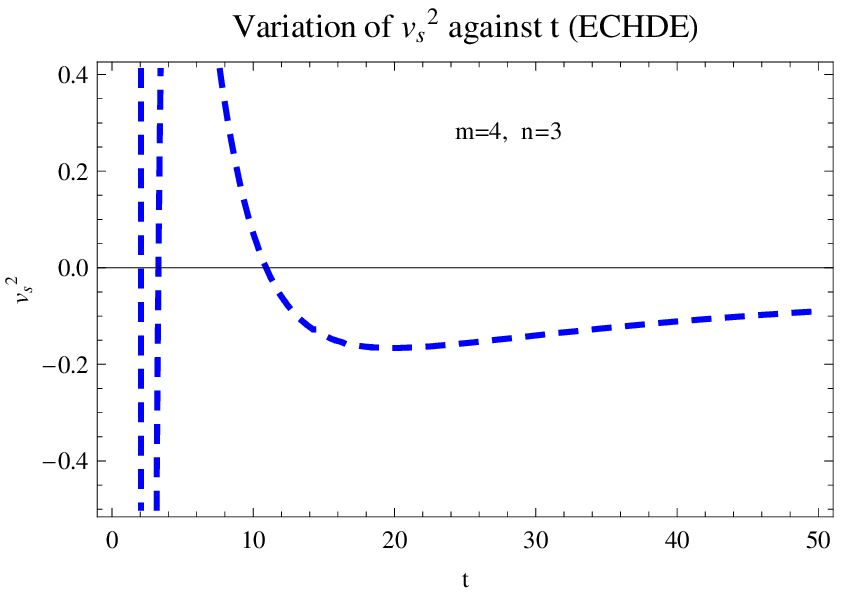}~~~~~~~~~\includegraphics[height=3in, width=3in]{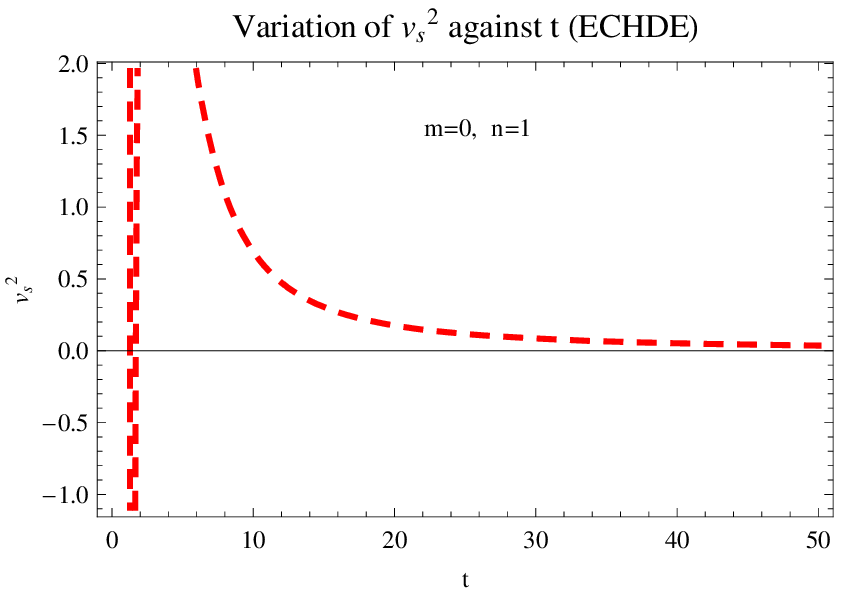}~~~~~~~\\
\vspace{1mm}
~~~~~~~~~~~~~~~~~~~~~~~~~~~~~~~~~~~Fig.5~~~~~~~~~~~~~~~~~~~~~~~~~~~~~~~~~~~~~~~~~~~~~~~~~~~~~~~~~~~~~~~Fig.6~~~~~~~~~~~~~~~~~\\
\vspace{3mm} \textit{\textbf{Fig 5, 6} shows the variation of
squared speed of sound $v_{s}^{2}$ against time $t$ for ECHDE dark
energy for different values of $m$ and $n$.}

\end{figure}

\section{Conclusion}
Here we have discussed the reconstruction scheme of the $f(R)$
gravity theory with two different dark energy models, namely,
$(m,n)$-type holographic dark energy (HDE) and entropy corrected
$(m,n)$-type holographic dark energy (ECHDE). Power law form of
scale factor has been taken into account. Using the correspondence
between the gravity and dark energy we have calculated the
explicit forms of $f(R)$ in terms of $R$ for both the dark energy
models. To study the nature of the calculated models plots have
been generated. It is seen that for HDE case, the reconstructed
model grows with R, whereas for the ECHDE case, the model decays.
The result is in agreement with previous works found in literature
\cite{Rahul}. Moreover, from literature \cite{Huang} the best-fit
analysis indicates that the models with $m=n+1$ and relatively
small values of $m$ is more favoured. In our analysis we tried to
justify/injustify the above fact by adopting a value satisfying
the above rule and a value violating it. Our results justified the
rule as the parametric values satisfying the rule produced more
physically justificable results. Moreover the stability of the
reconstructed models have been brought under the scanner using the
squared speed of sound in each case for two different sets of
values of the parameters $m$ and $n$. In the HDE case it is seen
that for $m=4, n=3$, the model exhibits classical instability, but
for $m=0, n=1$ the model is perfectly stable. For ECHDE case, when
$m=4, n=3$, we witness a transition from stable to unstable state.
But for $m=0, n=1$, the model is absolutely stable.

\section*{Acknowledgements}

The author sincerely acknowledges the facilities provided by the
Inter-University centre for Astronomy and Astrophysics (IUCAA),
Pune, India where a part of the work was carried out. The author
also acknowledges the anonymous referee for enlightening comments
that helped to improve the quality of the manuscript.\\


\begin{thebibliography}{99}
\bibitem{Perlmutter} S. J. Perlmutter et al. :- {\it Nature} {\bf 391} 51 (1998).
\bibitem{Riess} A. G. Riess et al. [Supernova Search Team Collaboration] :- {\it Astron. J.} {\bf 116} 1009 (1998).
\bibitem{Spergel} D. N. Spergel et al. :- {\it Astrophys. J. Suppl. Ser.} {\bf 170} 377 (2007).
\bibitem{Bamba1} K. Bamba et al. :- {\it Phys. Lett. B} {\bf 679} 282 (2009).
\bibitem{Bamba2} K. Bamba et al. :- {\it Phys. Rev. D} {\bf 79} 083014 (2009).
\bibitem{Nojiri1} S. Nojiri, S. D. Odintsov :- {\it Phys. Rev. D} {\bf 74} 086005 (2006).
\bibitem{Bennett} SDSS Collaboration (C.L. Bennet et al.) :- {\it Astrophys. J.} {\bf 583} 1 (2003).
\bibitem{Fisher1} W. Fischler et al. :- {\it arXiv:hep-th/9806039} (1998).
\bibitem{Susskind1} L. Susskind, :- {\it J. Math. Phys.} {\bf 36} 6377 (1995).
\bibitem{Cohen1} A. G. Cohen et al. :- {\it Phys. Rev. Lett.} {\bf 82} 4971 (1999).
\bibitem{Li1} M. Li :- {\it Phys. Lett. B} {\bf 603} 1 (2004).
\bibitem{Pavon1} D. Pavon et al. :- {\it Phys. Lett. B} {\bf 628} 206 (2005).
\bibitem{Wang1} B. Wang et al. :- {\it Phys. Lett. B} {\bf 624} 141 (2005).
\bibitem{Wei1} H. Wei et al. :- {\it Phys. Lett. B} {\bf 663} 1 (2008).
\bibitem{Granda1} L. N. Granda et al. :- {\it Phys. Lett. B} {\bf 669} 275 (2008).
\bibitem{Setare1} M. R. Setare et al. :- {\it Europhys. Lett.} {\bf 92} 49003 (2010).
\bibitem{Wei2} H. Wei, R.-G. Cai :- {\it Phys. Lett. B} {\bf 660} 113 (2008).
\bibitem{Gao1} C. Gao et al. :- {\it Phys. Rev. D} {\bf 79} 043511 (2009).
\bibitem{Ling1} Y. Ling :- {\it Mod. Phys. Lett. A} {\bf 28} 1350128 (2013).
\bibitem{Sharif} M. Sharif et. al. :- {\it Eur. Phys. J. C} {\bf 72} 2097 (2012).
\bibitem{Kim} K. Kim et al. :- {\it Phys. Lett. B} {\bf 660} 118 (2008).
\bibitem{Myung} Y. S. Myung :- {\it Phys. Lett. B} {\bf 652} 223 (2007).
\bibitem{Rahul} R. Ghosh, U. Debnath :- {\it Eur. Phys. J. Plus} {\bf 129} 81 (2014).
\bibitem{Huang} Z.-P. Huang, Y.-L. Wu :- {\it JCAP} {\bf 07} 035 (2012) {\it arXiv:1205.0608 [gr-qc]}.
\bibitem{Capo} S. Capozziello, V. F. Cardone, S. Carloni, A. Troisi :- {\it Int. J. Mod. Phys. D} {\bf 12} 1969 (2003).
\bibitem{Nojiri2} S. Nojiri, S. D. Odintsov :- {\it Gen. Relativ. Gravit.} {\bf 36} 1765 (2003).
\bibitem{Staro} A. A. Starobinski :- {\it Phys. Lett. B} {\bf 91} 99 (1980).
\bibitem{Soti} T. P. Sotiriou, V. Faraoni :- {\it Rev. Mod. Phys.} {\bf 82} 451 (2010).
\bibitem{Felice} A. D. Felice, S. Tsujikawa :- {\it Living Rev. Rel.} {\bf 13} 3 (2010).
\bibitem{Cognola} G. Cognola, E. Elizalde, S. Nojiri, S. D. Odintsov, L. Sebastiani, S. Zerbini :- {\it Phys. Rev. D} {\bf 77} 046009 (2009)
\bibitem{Elizalde} E. Elizalde, S. Nojiri, S. Odintsov, L. Sebastiani, S. Zerbini :- {\it Phys. Rev. D} {\bf 83} 086006 (2011).
\bibitem{Nojiri3} S. Nojiri, S. D. Odintsov :- {\it J. Phys. A} {\bf 40} 6725 (2007)

\end{thebibliography}
\end{document}